\documentclass[11pt]{article}
\usepackage{amssymb,amsmath,amsfonts}
\usepackage{graphicx}
\usepackage{graphics}
\usepackage{eepic,epsfig}

\begin{document}

\title{ Photoproduction of electron-positron pairs in the presence of
hyperacoustic oscillations}
\author{A. R. Mkrtchyan, L. Sh. Grigoryan, A. A. Saharian\thanks{%
E-mail: saharyan@server.physdep.r.am}, V. V. Parazian \\
%EndAName
\textit{\small Institute of Applied Problems in Physics, 25 Nersessian Str.,
375014 Yerevan, Armenia}}
\maketitle

\begin{abstract}
We report on the recent progress in the investigation of the influence of
hyperacoustic vibrations on the coherent electron-positron pair creation by
high-energy photons in crystals. In dependence of the values for the
parameters, the presence of the deformation field can either enhance or
reduce the cross-section. This can be used to control the parameters of the
positron sources for storage rings and colliders.
\end{abstract}


\section{Introduction}

\label{sec1}

The basic source to creating positrons for high-energy electron-positron
colliders is the electron-positron pair creation by hard bremsstrahlung
photons produced when a powerful electron beam hits an amorphous target. One
possible approach to increase the positron production efficiency is to use a
crystal target as a positron emitter. If the formation length exceeds the
interatomic spacing, the interference effects from all atoms within this
length are important and the cross-section for the pair creation can change
essentially compared with the corresponding quantities for a single atom.
From the point of view of controlling the parameters of various processes in
a medium, it is of interest to investigate the influence of external fields,
such as acoustic waves, temperature gradient etc., on the corresponding
characteristics. Our considerations of diffraction radiation, transition
radiation, parametric X-radiation, channelling radiation, bremsstrahlung by
high-energy electrons, have shown that the external fields can essentially
change the angular-frequency characteristics of the radiation intensities.
In \cite{Mkrt03,Mkrt05a,Mkrt05b} we have investigated the influence of the
hypersonic wave excited in a crystal on the process of electron-positron
pair creation by high-energy photons. The case of simplest crystal with one
atom in the lattice base and the sinusoidal deformation field generated by
the hypersound were considered in \cite{Mkrt03}. To have a considerable
influence of the acoustic wave, high-frequency hypersound is needed. Usually
this type of waves is excited by high-frequency electromagnetic field
through the piezoelectric effect in crystals with a complex lattice base. In
\cite{Mkrt05a} we have generalized the results of \cite{Mkrt03} for crystals
with a complex base and for acoustic waves with an arbitrary profile. In
particular, the numerical calculations are carried out for the quartz single
crystal and for the photons of energy 100 GeV. The results of the numerical
calculations on the base of the formulae given in \cite{Mkrt05a} for the
pair creation cross-section by the photons of energy 3.5 GeV are presented
in the recent paper \cite{Mkrt05b}, where the scheme of experimental setup
is proposed for the corresponding measurements on the photon channel of the
Yerevan synchrotron. The influence of acoustic wanes to the bremsstrahlung
by high-energy electrons in crystals is investigated in \cite{Mkrtbrem}.

\section{Cross-section for the coherent pair creation}

\label{sec2}

\begin{figure}[tbph]
\begin{center}
\epsfig{figure=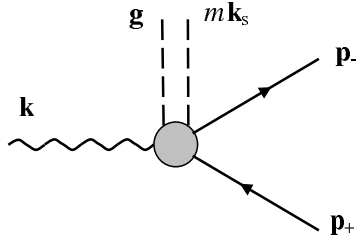,width=5cm,height=3.5cm}
\end{center}
\caption{The pair creation by high-energy
photon with the momentum ${\mathbf{k}}$. Vectors ${\mathbf{g}}$ and $m{\mathbf{k}}%
_{s}$ stand for the momenta transferred to the crystal and to the external
field.}
\label{fig1}
\end{figure}

Consider the creation of electron-positron pairs by high-energy photons in a
crystal (see figure \ref{fig1} for the diagram of the process). We denote by
$(\omega ,\mathbf{k})$, $(E_{+},\mathbf{p}_{+})$, and $(E_{-},\mathbf{p}_{-})
$ the energies and momenta for the photon, positron, and electron
respectively. When acoustic waves are excited the positions of atoms in the
crystal can be written as $\mathbf{r}_{n0}^{(j)}=\mathbf{r}_{ne}^{(j)}+%
\mathbf{u}_{n}^{(j)}$, where $\mathbf{r}_{ne}^{(j)}$\ determines the
equilibrium position of an atom in the situation without the deformation, $%
\mathbf{u}_{n}^{(j)}$ is the displacement of the atom caused by the acoustic
field. The collective index $n$ enumerates the elementary cell and the
superscript $j$ enumerates the atoms in a given cell of a crystal. We will
consider displacements of the form%
\begin{equation}
\mathbf{u}_{n}^{(j)}=\mathbf{u}_{0}f(\mathbf{k}_{s}\mathbf{r}_{ne}^{(j)}),
\label{un}
\end{equation}%
where $\mathbf{u}_{0}$ and $\mathbf{k}_{s}$\ are the amplitude and wave
vector of the acoustic wave, $f(x)$ is an arbitrary function with the period
$2\pi $, $\max f(x)=1$. For a lattice with a complex cell the coordinates of
the atoms can be presented in the form $\mathbf{r}_{ne}^{(j)}=\mathbf{R}_{n}+%
\mathbf{\rho }^{(j)}$, where $\mathbf{R}_{n}$\ determines the positions of
the atoms for one of primitive lattices, and $\mathbf{\rho }^{(j)}$\ are the
equilibrium positions for other atoms inside $n$-th elementary cell with
respect to $\mathbf{R}_{n}$.

The cross-section for the pair creation process is presented in the form $%
d\sigma =N_{0}(d\sigma _{n}+d\sigma _{c})$, where $d\sigma _{n}$ and $%
d\sigma _{c}$ are the incoherent and coherent parts of the cross-section per
atom and $N_{0}$ is the number of atoms in the crystal. The coherent part is
determined by the formula%
\begin{equation}
\frac{d\sigma _{c}}{dE_{+}}=\frac{e^{2}N}{\omega ^{2}N_{0}\Delta }\sum_{m,%
\mathbf{g}}\frac{g_{m\perp }^{2}}{g_{m\parallel }^{2}}\left( \frac{\omega
^{2}}{2E_{+}E_{-}}-1+\frac{2\delta }{g_{m\parallel }}-\frac{2\delta ^{2}}{%
g_{m\parallel }^{2}}\right) \left\vert F_{m}(\mathbf{g}_{m}\mathbf{u}%
_{0})\right\vert ^{2}\left\vert S(\mathbf{g}_{m},\mathbf{g})\right\vert ^{2},
\label{dsigmac/dE+}
\end{equation}%
where $e$ is the electron charge, $N$ is the number of cells in the crystal,
$\Delta =a_{1}a_{2}a_{3}$ is the unit cell volume for an orthogonal lattice
with the lattice constants $a_{1}$, $a_{2}$, $a_{3}$,
\begin{equation}
\mathbf{g}_{m}=\mathbf{g}-m\mathbf{k}_{s},\;m=0,\pm 1,\pm 2,\ldots ,
\label{gmn}
\end{equation}%
$\mathbf{g}$ is the reciprocal lattice vector, $\mathbf{g}_{m\parallel }$
and $\mathbf{g}_{m\perp }$ are the parallel and perpendicular components of
the vector $\mathbf{g}_{m}$ with respect to the direction of the photon
momentum $\mathbf{k}$, $\delta =1/l_{c}$ is the minimum longitudinal
momentum transfer, and $l_{c}=2E_{+}E_{-}/(m_{e}^{2}\omega )$ is the
formation length for the pair creation process. In (\ref{dsigmac/dE+}) the
summation goes under the constraint $g_{m\parallel }\geq \delta $. The
function $F_{m}(x)$\ is the Fourier-transform of the function $e^{ixf(t)}$:
\begin{equation}
F_{m}(x)=\frac{1}{2\pi }\int_{-\pi }^{\pi }e^{ixf(t)-imt}dt,  \label{Fm}
\end{equation}%
and $S(\mathbf{g}_{m},\mathbf{g})=\sum_{j}u_{\mathbf{g}_{m}}^{(j)}e^{i%
\mathbf{g\rho }^{(j)}}e^{-\frac{1}{2}g_{m}^{2}\overline{u_{t}^{(j)2}}}$ is
the factor determined by the structure of the crystal lattice base, $%
\overline{u_{t}^{(j)2}}$ is the temperature dependent mean-squared amplitude
of the thermal vibrations of the $j$-th atom. The corresponding momentum
conservation is written in the form%
\begin{equation}
\mathbf{k}=\mathbf{p}_{+}+\mathbf{p}_{-}+\mathbf{g}-m\mathbf{k}_{s},
\label{momcons}
\end{equation}%
where $-m\mathbf{k}_{s}$ stands for the momentum transfer to the external
field. Formula (\ref{dsigmac/dE+}) differs from the formula in an undeformed
crystal by the replacement $\mathbf{g}\rightarrow \mathbf{g}_{m}$, and by an
additional summation over $m$ with weights $\left\vert F_{m}(\mathbf{g}_{m}%
\mathbf{u}_{0})\right\vert ^{2}$. This corresponds to the presence of an
additional one-dimensional superlattice with the period $\lambda _{s}=2\pi
/k_{s}$ and the reciprocal lattice vector $m\mathbf{k}_{s}$. As the main
contribution into the cross-section comes from the terms with $g_{m\parallel
}\sim \delta $, the influence of the deformation field may be considerable
if $|mk_{s\parallel }|\gtrsim \delta $. Combining this with the estimate
that the main contribution comes from the terms for which $\left\vert m%
\mathbf{k}_{s}\mathbf{u}_{0}\right\vert \lesssim \left\vert \mathbf{gu}%
_{0}\right\vert $, or equivalently $\left\vert m\right\vert \lesssim \lambda
_{s}/a$, we find the condition: $u_{0}/\lambda _{s}\gtrsim a/4\pi ^{2}l_{c}$%
. At high energies one has $a/l_{c}\ll 1$ and this condition can be
consistent with the condition $u_{0}/\lambda _{s}\ll 1$.

\begin{figure}[tbph]
\begin{center}
\epsfig{figure=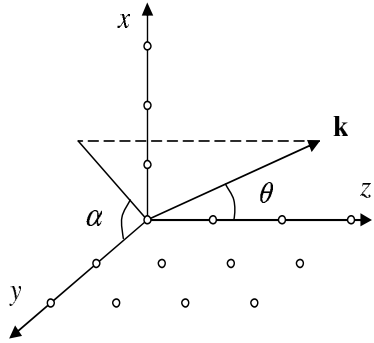,width=5cm,height=4.5 cm}
\end{center}
\caption{Geometry of the problem in the case of an orthogonal
lattice with the axes $(x,y,z)$} \label{fig2}
\end{figure}

If the photon moves in a non-oriented crystal, in formula (\ref{dsigmac/dE+}%
) the summation over $\mathbf{g}$\ can be replaced by the integration and
the pair creation cross-section coincides with that in an amorphous medium.
The role of coherence effects in the pair creation cross-section is
essential when the photon enters into the crystal at small angle $\theta $
with respect to a crystallographic axis (axis $z$ in our consideration). In
this case the main contribution into the coherent part of the cross-section
comes from the crystallographic planes, parallel to the chosen axis which
correspond to the summands with $g_{z}=0$. The behavior of this
cross-section as a function on the positron energy essentially depends on
the angle $\alpha $ between the projection of the photon momentum on the
plane $(x,y)$ and $y$-axis (see figure \ref{fig2}).
If the photon moves far from the corresponding crystallographic planes, the
summation over the perpendicular components of the reciprocal lattice vector
can be replaced by the integration: $\sum_{g_{x},g_{y}}\rightarrow \left[
a_{1}a_{2}/(2\pi )^{2}\right] \int dg_{x}dg_{y}$. When the photon enters
into the crystal near a crystallographic plane ($\alpha $ is small), two
cases have to be distinguished. For the first one $\theta \sim a_{2}/2\pi
l_{c}$, the summation over $g_{x}$ can be replaced by integration. Under the
assumption $\mathbf{u}_{0}\perp \mathbf{a}_{1}$, the corresponding formula
is further simplified to the form%
\begin{equation}
\frac{d\sigma _{c}}{dE_{+}}\approx \frac{e^{2}N\omega ^{-2}}{2\pi
N_{0}a_{2}a_{3}}\sum_{m,g_{y}}\frac{\left\vert F_{m}(g_{y}u_{0y})\right\vert
^{2}}{g_{m\parallel }^{2}}\left( \frac{\omega ^{2}}{2E_{+}E_{-}}-1+\frac{%
2\delta }{g_{m\parallel }}-\frac{2\delta ^{2}}{g_{m\parallel }^{2}}\right)
\int dg_{x}g_{\perp }^{2}\left\vert S(\mathbf{g}_{m},\mathbf{g})\right\vert
^{2},  \label{dsigmac35/dE+}
\end{equation}
with an effective structure factor determined by the integral on the right
and $g_{m\parallel }\approx -mk_{z}+\theta g_{y}\geq \delta $. In the second
case we assume that $\delta \sim 2\pi \theta \alpha /a_{1}$. Now the main
contribution comes from the terms with $g_{y}=0$ and the formula for the
cross-section takes the form%
\begin{equation}
\frac{d\sigma _{c}}{dE_{+}}\approx \frac{e^{2}N}{\omega ^{2}N_{0}\Delta }%
\sum_{m,g_{x}}\frac{g_{\perp }^{2}}{g_{m\parallel }^{2}}\left( \frac{\omega
^{2}}{2E_{+}E_{-}}-1+\frac{2\delta }{g_{m\parallel }}-\frac{2\delta ^{2}}{%
g_{m\parallel }^{2}}\right) \left\vert F_{m}(\mathbf{g}_{m}\mathbf{u}%
_{0})\right\vert ^{2}\left\vert S(\mathbf{g}_{m},\mathbf{g})\right\vert ^{2},
\label{dsigmac4/dE+}
\end{equation}%
where $g_{m\parallel }\approx -mk_{z}+\psi g_{x}$, $\psi =\alpha \theta $,
and the summation goes under the condition $g_{m\parallel }\geq \delta $.

The numerical calculations for the cross-section are carried out in the case
of $\mathrm{SiO}_{2}$ single crystal with the Moliere parametrization of the
atomic potentials and for the deformation field generated by the transversal
sinusoidal acoustic wave of $S$ type with frequency $5$ GHz. To deal with an
orthogonal lattice, we choose as an elementary cell the cell including 6
atoms of silicon and 12 atoms of oxygen (Shrauf elementary cell). For this
choice the $y$ and $z$ axes of the orthogonal coordinate system $(x,y,z)$
coincide with the standard $Y$ and $Z$ axes of the quartz crystal, whereas
the angle between the axes $x$ and $X$ is equal to $\pi /6$. The vector of
the amplitude of the displacement is directed along $x$-direction, $\mathbf{u%
}_{0}=(u_{0},0,0)$, and the velocity is $4.687\cdot 10^{5}$ cm/sec. The
vector determining the direction of the hypersound propagation lies in the
plane $YZ$ and has the angle with the axis $Z$ equal to $0.295$ rad. The
numerical calculations for various values of the parameters in the problem
show that, in dependence of the values for the parameters, the presence of
the deformation field can either enhance or reduce the cross-section. This
can be used to control the parameters of the positron sources for storage
rings and colliders.

\begin{figure}[tbph]
\begin{center}
\begin{tabular}{ccc}
\epsfig{figure=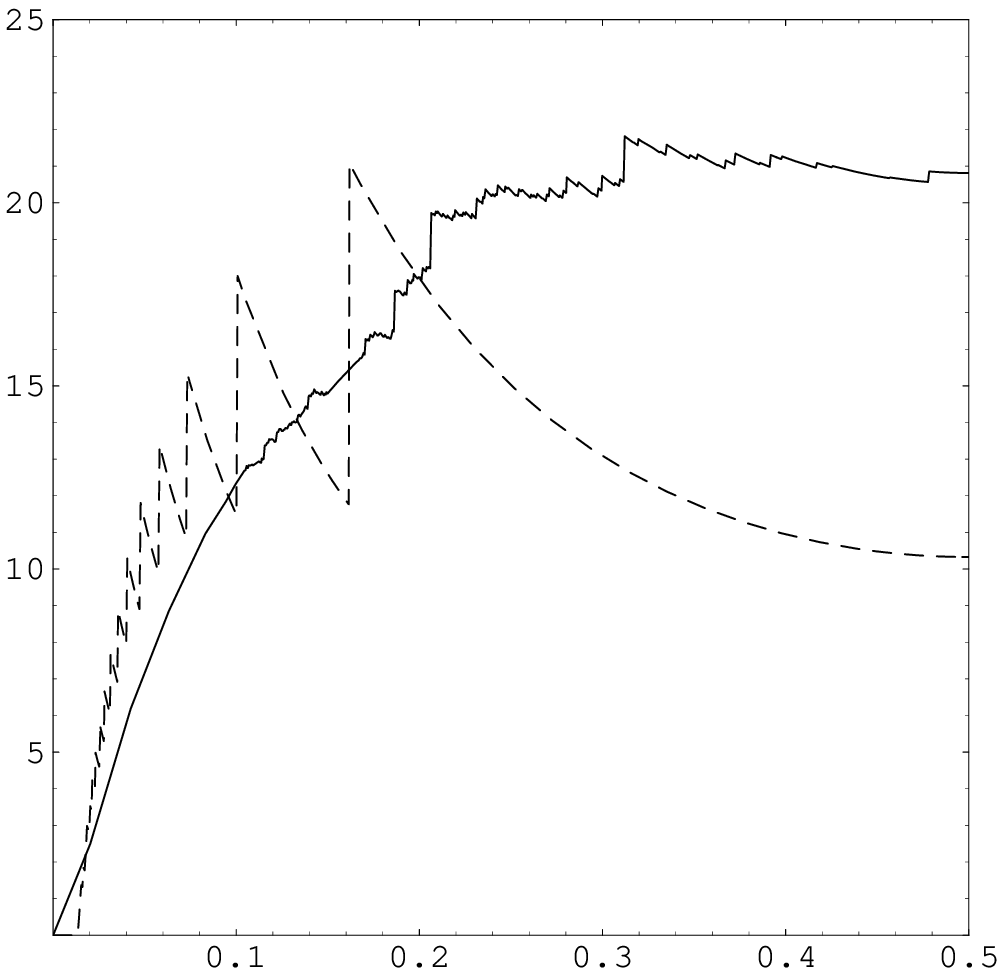,width=5cm,height=5cm} & \hspace*{0.5cm} & %
\epsfig{figure=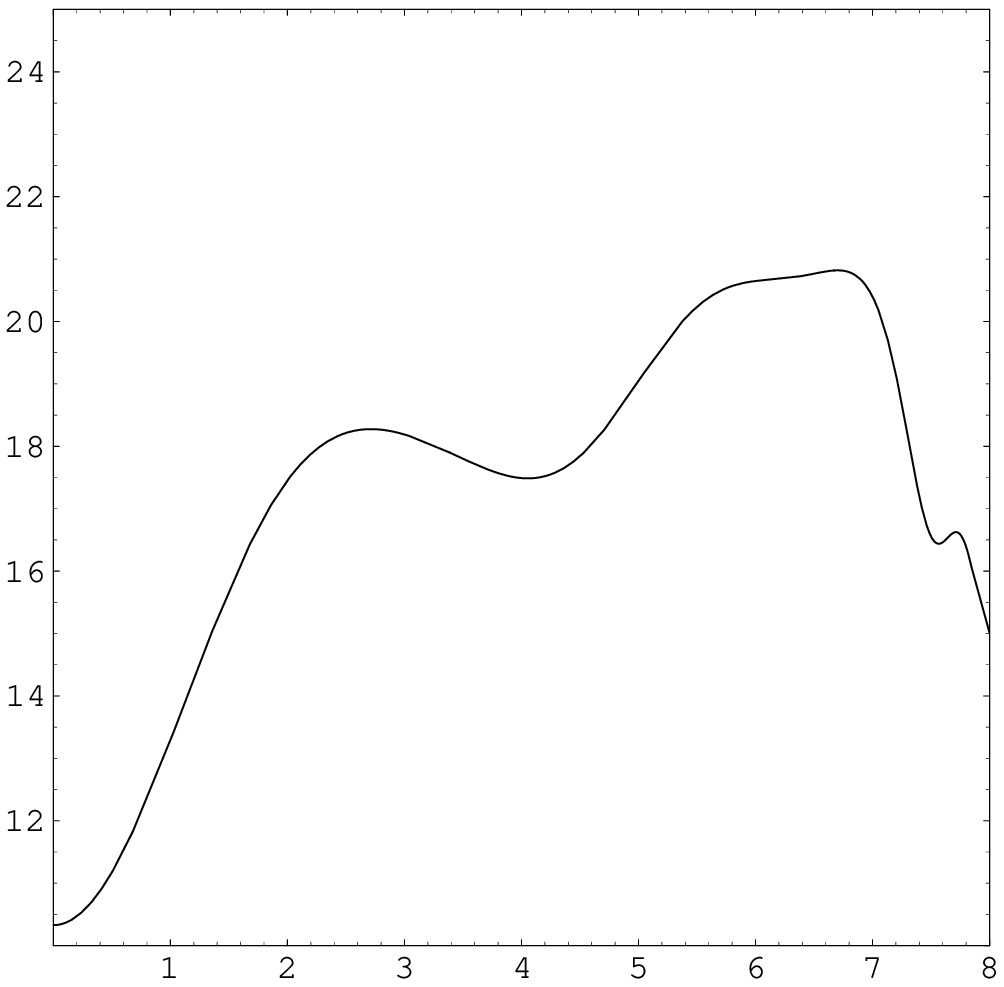,width=5cm,height=5cm}%
\end{tabular}%
\end{center}
\caption{Coherent part of the cross-section, $10^{-3}(m_{e}^{2}%
\protect\omega /e^{6})d\protect\sigma _{c}/dE_{+}$, in the quartz single
crystal for the sinusoidal transversal acoustic wave of the $S$ type with
frequency 5 GHz, evaluated by formula (\protect\ref{dsigmac35/dE+}), as a
function of $E_{+}/\protect\omega $ (left panel) for $u_{0}=0$ (dashed
curve), $2\protect\pi u_{0}/a_{2}=6.75$ (full curve) and as a function of $2%
\protect\pi u_{0}/a_{2}$ (right panel) for the positron energy corresponding to $E_{+}/%
\protect\omega =0.5$. The values for the other parameters are as follows: $%
\protect\theta =0.0042$ rad, $\protect\omega =50$ GeV.}
\label{fig3}
\end{figure}

As an illustration of the enhancement in figure \ref{fig3} (left panel) we
have depicted the quantity $d\sigma _{c}/dE_{+}$ evaluated by formula (\ref%
{dsigmac35/dE+}) as a function of the ratio $E_{+}/\omega $ for $u_{0}=0$
(dashed curve) and $2\pi u_{0}/a_{2}=6.75$ (full curve). The deformation is
induced by the transversal acoustic wave of the $S$ type with frequency 5
MHz and $\theta =0.0042$ rad, $\omega =50$ GeV. As the cross-section is
symmetric under the replacement $E_{+}/\omega \rightarrow 1-E_{+}/\omega $,
we have plotted the graphs for the region $0\leq E_{+}/\omega \leq 0.5$
only. In figure \ref{fig3} (right panel) the cross-section evaluated by
formula (\ref{dsigmac35/dE+}) is presented as a function of $2\pi u_{0}/a_{2}
$ for the positron energy corresponding to $E_{+}/\omega =0.5$. The values
of the other parameters are the same as those for the left panel. Note that
for the chosen values of the parameters one has $\lambda _{s}\approx
9.4\times 10^{-4}$cm, whereas $l_{c}\approx 1.9\times 10^{-6}$cm for the
energies $E_{+}=E_{-}=25$ Gev and, hence, $\lambda _{s}\gg l_{c}$.

\begin{figure}[tbph]
\begin{center}
\begin{tabular}{ccc}
\epsfig{figure=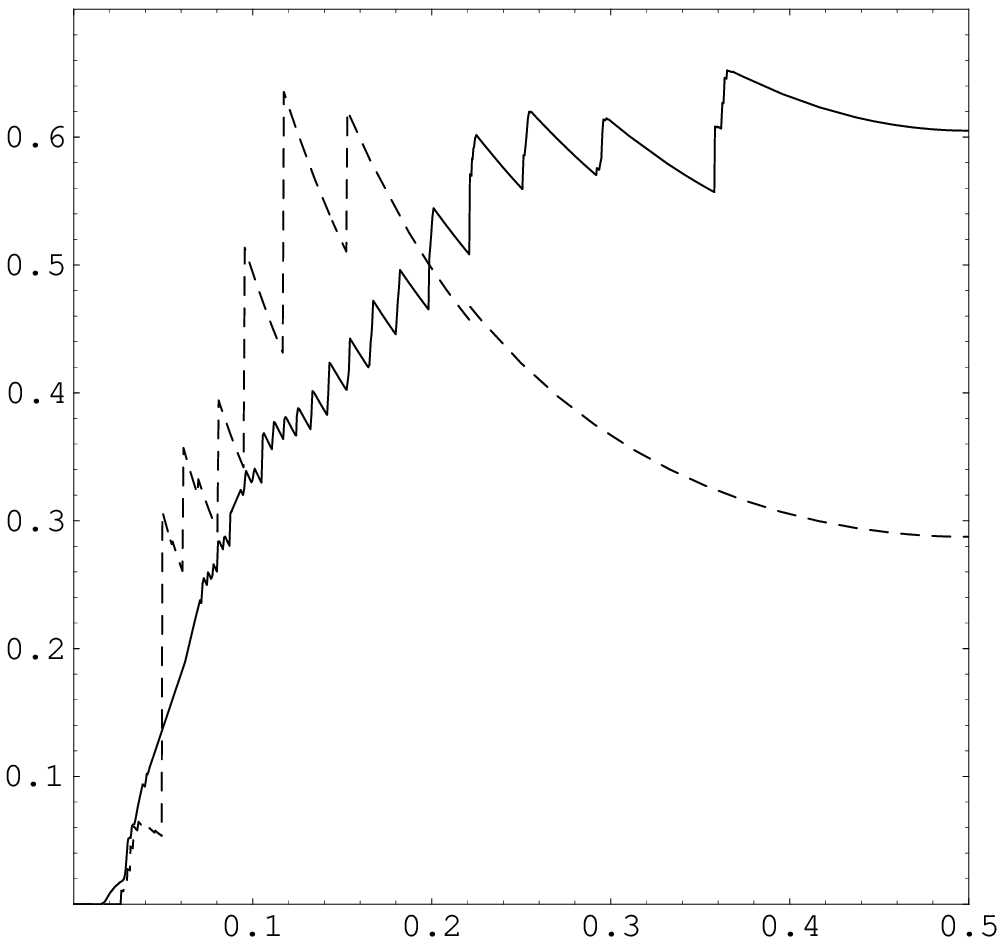,width=5cm,height=5cm} & \hspace*{0.5cm} & %
\epsfig{figure=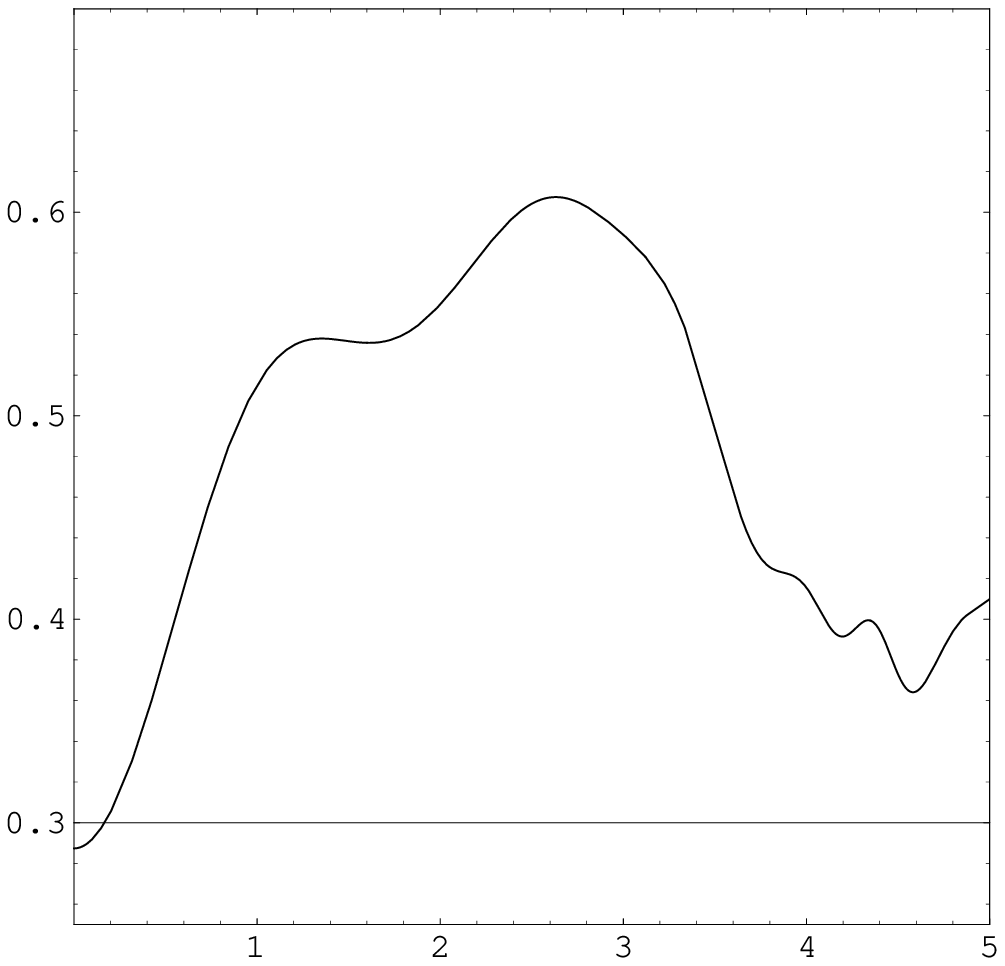,width=5cm,height=5cm}%
\end{tabular}%
\end{center}
\caption{Pair creation cross-section, $10^{-3}(m_{e}^{2}\protect\omega %
/e^{6})d\protect\sigma _{c}/dE_{+}$, evaluated by formula (\protect\ref%
{dsigmac4/dE+}), as a function of $E_{+}/\protect\omega $ (left panel) for $%
u_{0}=0$ (dashed curve), $2\protect\pi u_{0}/a_{2}=2.75$ (full curve) and as
a function of $2\protect\pi u_{0}/a_{2}$ (right panel) for the positron
energy corresponding to $E_{+}/\protect\omega =0.5$. The values for the
other parameters are as follows: $\protect\psi =0.0022$, $\protect\omega %
=50$ GeV.}
\label{fig4}
\end{figure}

In figure \ref{fig4} (left panel) we have presented the cross-section
evaluated by formula (\ref{dsigmac4/dE+}) as a function of the ratio $%
E_{+}/\omega $ for $u_{0}=0$ (dashed curve) and $2\pi u_{0}/a_{1}=2.75$
(full curve) in the case $\psi =0.0022$. The values for the other parameters
are the same as in figure \ref{fig3}. In figure \ref{fig4} (right panel) we
have plotted the cross-section evaluated by formula (\ref{dsigmac4/dE+}) as
a function of $2\pi u_{0}/a_{1}$ for the positron energy corresponding to $%
E_{+}/\omega =0.5$ and for $\psi =0.001$. The values for the other
parameters are the same as for the left panel.

\end{document}